\documentclass[sigconf]{acmart}
\AtBeginDocument{%
  }

\copyrightyear{2026}
\acmYear{2026}
\setcopyright{cc}
\setcctype{by}
\acmConference[WebSci '26]{18th ACM Web Science Conference}{May 26--29, 2026}{Braunschweig, Germany}
\acmBooktitle{18th ACM Web Science Conference (WebSci '26), May 26--29, 2026, Braunschweig, Germany}
\acmDOI{10.1145/3795766.3799759}
\acmISBN{979-8-4007-2504-3/2026/05}

\usepackage{tabularx}
\begin{document}

\title{From Guidelines to Practice: Evaluating the Reproducibility of Methods in Computational Social Science}

\author{Fakhri Momeni}
\email{fakhri.momeni@gesis.org}
\orcid{0000-0002-5572-575X}
\author{Sarah Sajid}
\email{sarah.sajid@gesis.org}
\orcid{}
\author{Johannes Kiesel}
\orcid{0000-0002-1617-6508}
\email{Johannes.Kiesel@gesis.org}
\affiliation{%
  \institution{Gesis - Leibniz Institute for Social Sciences}
  \city{Cologne}
  \country{Germany}
}


\begin{abstract}
Reproducibility remains a central challenge in computational social science, where complex workflows, evolving software ecosystems, and inconsistent documentation hinder researchers’ ability to re-execute published methods. This study presents a systematic evaluation of reproducibility across three conditions: uncurated documentation, curated documentation, and curated documentation paired with a preset execution environment. Using 47 usability test sessions, we combine behavioral performance indicators (success rates, task time, and error profiles) with questionnaire data and thematic analysis to identify technical and conceptual barriers to reproducibility.

Curated documentation substantially reduced repository-level errors and improved users’ ability to interpret method outputs. Standardizing the execution environment further improved reproducibility, yielding the highest success rate and shortest task completion times. Across conditions, participants frequently relied on AI tools for troubleshooting, often enabling independent resolution of issues without facilitator intervention.

Our findings demonstrate that reproducibility barriers are multi-layered and require coordinated improvements in documentation quality, environment stability, and conceptual clarity. We discuss implications for the design of reproducibility platforms and infrastructure in computational social science.
\end{abstract}



\maketitle

\section{Introduction}

Reproducibility is widely recognized as a cornerstone of scientific progress, yet it remains a persistent challenge across computational fields, including machine learning, natural language processing, and computational social science \cite{pineau2021improving, belz2021systematic, magnusson2023reproducibility}. As research increasingly relies on complex software pipelines, heterogeneous data sources, and rapidly evolving libraries, the ability to re-execute published methods has become both more important and more difficult to ensure. Recent initiatives—such as reproducibility checklists \cite{magnusson2023reproducibility, Momeni2025ChecklistsSocialScience}, replication packages \cite{chao2020replicability}, and workflow standardization efforts—have highlighted the need for clearer documentation and more robust execution environments, but empirical evidence on their effectiveness for end users remains limited.

Computational social science presents distinct reproducibility challenges due to its reliance on diverse analytical methods, varied programming languages, and installation-sensitive tools \cite{weller2016manifesto, hutton2015making}. Inadequate documentation, unstable dependencies, and incomplete examples frequently impede users’ ability to reproduce results, even when code is publicly available \cite{chen2022reproducibility, gundersen2018state}. Prior work has emphasized the importance of structured documentation and metadata, but comparatively little is known about how different forms of documentation and environment support influence users’ actual reproduction success, error patterns, and interpretive confidence.

This study addresses this gap by conducting a systematic evaluation of reproducibility under three conditions: uncurated documentation, curated documentation, and curated documentation paired with a preset execution environment. Through 47 usability tests, we examine not only whether users can reproduce computational social science methods, but also \emph{how} and \emph{why} reproduction succeeds or fails. We integrate behavioral KPIs (``key performance indicators'', specifically success rates, time on task, and error distributions) with post-test questionnaire responses and thematic analysis of participant reflections to identify the technical, procedural, and conceptual factors that shape reproducibility outcomes.

This paper makes the following contributions:
\begin{enumerate}
    \item An empirical comparison of three reproduction conditions, demonstrating the effect of documentation quality and environment standardization in supporting reproducibility.
    \item A detailed error analyses showing how curated documentation shifts failure modes from technical breakdowns to conceptual misunderstandings, while a standardized execution environment eliminates nearly all system-level errors.
    \item An empirical account of how participants used AI-assisted troubleshooting during reproduction attempts, highlighting its emerging role in practical reproducibility workflows.
\end{enumerate}

Together, these contributions advance our understanding of how technical and conceptual support for the reproducibility of computational social science methods interact, and they provide actionable guidance for designing tools, documentation practices, and environments that better support transparent and verifiable research.

\section{Related Work}

\subsection{Reproducibility Issues in Machine Learning and NLP}

Reproducibility is increasingly recognised as a core requirement for trustworthy ML and NLP research, yet a substantial body of work documents a persistent ``reproducibility crisis'' in these fields \cite{pineau2021improving,belz2021quantifying,magnusson2023reproducibility,belz2022metrological,belz2021systematic,beam2020challenges,mcdermott2021reproducibility,nair2024exploring,semmelrock2025reproducibility}. Quantitative assessments show that only a small fraction of reported results can be exactly reproduced: one large-scale analysis reports that just 14.03\% of score pairs match exactly, and in 59.2\% of cases reproduction attempts yield worse performance than originally reported \cite{chen2022reproducibility}. In parallel, only around half of papers open-source their code \cite{magnusson2023reproducibility}, and many omit key implementation details, hyperparameter settings, or uncertainty estimates \cite{chen2022reproducibility,casola2022pre}.

Several strands of work disentangle the underlying sources of non-reproducibility. First, there is considerable \emph{terminological ambiguity}. Different communities use terms such as reproducibility and replicability inconsistently, often distinguishing between exact reruns with the same code and data (technical or R1 reproducibility) and conceptual replication with different data \cite{pineau2021improving,belz2021quantifying,belz2022metrological,nair2024exploring,semmelrock2025reproducibility}. Even within the narrow R1 setting, exact reproduction is often impossible due to technical and methodological factors.

Second, ML systems—especially deep learning models—exhibit strong \emph{computational and environmental sensitivity}. Stochastic optimisation, random initialisation, and non-deterministic hardware behaviour mean that unreported or uncontrolled random seeds can substantially change reported performance \cite{pineau2021improving,bouthillier2019unreproducible,ahmed2022managing,albertoni2023reproducibility,semmelrock2023reproducibility}. Small differences in hardware, library, or framework versions can also break pipelines or alter outcomes \cite{lynnerup2020survey,shahriari2022deep,jezequel2015estimation}. For very large models, such as BERT or contemporary LLMs, the computational cost of full reproduction can be prohibitive, with estimated costs in the order of millions of dollars \cite{raff2019step,beam2020challenges}.

Third, a range of \emph{methodological flaws} complicate trust in reported results. Data leakage is a prominent failure mode: spurious correlations arising from data handling, sampling, or preprocessing can inflate performance and collapse under realistic conditions \cite{kapoor2022leakage,kapoor2023leakage,kuhn2013applied}. Surveys of deep learning studies report low rates of accessible replication packages and sparse reporting of multiple runs or statistical significance \cite{chao2020replicability,casola2022pre}. Many models show instability under minor changes to vocabulary, dataset size, or training conditions \cite{chao2020replicability,chen2022reproducibility}, which challenges the robustness of conclusions.

Finally, NLP research that relies on \emph{human evaluation} faces additional reproducibility constraints. Experimental designs, annotation guidelines, and rating procedures are often insufficiently documented, leading to extremely low repeatability; one study estimates that only about 5\% of human evaluations are repeatable based solely on public information \cite{belz2023non}. Repeated data collection efforts for supervised tasks have shown consequential differences in labels and downstream model performance \cite{nair2024exploring}, particularly for cognitively complex evaluation dimensions \cite{belz2023non}.

Proposed interventions include reproducibility checklists, replication programmes at major conferences, and improved tooling for workflow management \cite{pineau2021improving,magnusson2023reproducibility,digan2021can}. Open-sourcing code is associated with higher reproducibility scores and reviewer confidence \cite{magnusson2023reproducibility}, but many suggested solutions remain weakly evaluated in practice \cite{magnusson2023reproducibility}. Overall, the literature highlights deep entanglements between technical environments, documentation practices, and evaluation protocols in ML and NLP.

\subsection{Reproducibility Challenges in Social Data Analysis and Computational Social Science}

Reproducibility is equally central in social data analysis and computational social science, where it underpins the validity and cumulative nature of empirical findings \cite{stodden2010data,weller2016manifesto,hutton2015making,hutton2015toward}. However, research based on social media and other digital behavioural data faces additional structural barriers.

A core challenge is \emph{data access and sharing}. Empirical studies report that only a small share of published work makes datasets available or citable—around 6.1\% in some reviews \cite{hutton2015making,hutton2015toward}. Access to platform data is often mediated by terms of service, proprietary APIs, and rate limits, creating a ``digital divide'' in which well-resourced or well-connected teams can obtain data that others cannot \cite{weller2016manifesto,mannheimer2017sharing}. Platforms such as Twitter typically restrict redistribution to identifiers (e.g., Tweet IDs), rather than full content, and may change or retire APIs over time, rendering legacy code unusable and hindering retrospective reproduction \cite{weller2016manifesto,mannheimer2017sharing,assenmacher2022benchmarking}.

Beyond access, \emph{documentation and metadata} are frequently insufficient. Social media studies employ heterogeneous data collection strategies, preprocessing pipelines, and analytical approaches, but rarely provide enough detail for independent researchers to reconstruct these choices \cite{hutton2015making,weller2016manifesto,morstatter2012opening}. High analytic flexibility—multiple reasonable choices at each processing step—creates a large researcher degrees-of-freedom problem, which can lead to substantial variability in results and selective reporting \cite{lazar2023functional,hutton2015making}. Without clear metadata for data collection, sampling, and cleaning, it becomes difficult to assess whether non-replication reflects methodological issues, platform drift, or genuine differences in behaviour.

Finally, reproducibility in social data analysis is constrained by \emph{ethical and legal considerations}. Sharing raw or richly annotated social media data raises serious privacy and re-identification risks, especially when combined with external data sources \cite{weller2016manifesto,morstatter2012opening,mannheimer2017sharing}. This tension between transparency and confidentiality complicates simple prescriptions for ``open data'' and requires carefully designed compromises in what is shared, how it is documented, and which components of a workflow can be made public.

\subsection{Checklists, Documentation, and README Files for Reproducibility}

Given these challenges, a growing literature investigates checklists, documentation practices, and README files as levers to improve computational reproducibility. Checklists have been proposed as formal mechanisms to structure reporting, reduce omissions, and establish expectations for authors and reviewers \cite{Du2022ChecklistMetabolomics,Kapoor2024REFORMS}. In NLP and ML, reproducibility checklists introduced at major conferences have been associated with improved reporting of efficiency, validation performance, hyperparameters, and summary statistics \cite{pineau2021improving,magnusson2023reproducibility}. For example, in the NeurIPS Reproducibility Program, over 75\% of submissions included reproducibility materials when prompted by a checklist, and higher checklist scores correlated with higher reviewer-assessed reproducibility and acceptance rates \cite{pineau2021improving,magnusson2023reproducibility}. Domain-specific checklists, such as REFORMS for ML-based scientific claims or metabolomics reporting checklists, similarly aim to prevent common errors and support reusable computational workflows \cite{Du2022ChecklistMetabolomics,Kapoor2024REFORMS}.

In parallel, a range of work emphasises the importance of \emph{documentation, metadata, and README files}. Recommendations include publishing precise code versions (e.g., via DOIs or commit hashes), reporting software and package versions, and documenting data source versions or access dates for dynamic resources \cite{Kapoor2024REFORMS,Olorisade2017TextMining,Feng2019ChecklistENM,tonzani2021star}. Empirical analyses of README files show that well-structured files with clear installation instructions, usage examples, and dependency information can lower technical barriers and improve information discoverability \cite{leipzig2021role,kim2018experimenting,prana2019categorizing}. At the same time, systematic gaps persist: many READMEs omit the purpose or status of the project, do not document assumptions, and cover only a fraction of the variables needed for full reproducibility \cite{prana2019categorizing,gundersen2018state}.

Crucially, most existing work evaluates documentation and checklists in terms of \emph{reporting quality}, not actual \emph{reproduction outcomes}. Tools have been developed to score README files against templates or classify their content \cite{akdeniz2023end,prana2019categorizing}, but they are rarely validated against empirical measures such as success rates, error types, or user experience in reproduction attempts \cite{akdeniz2023end,kim2018experimenting,leipzig2021role}. Case studies often find that reproduction remains difficult even when code and data are available, suggesting that documentation alone is necessary but not sufficient for reproducibility \cite{kim2018experimenting,leipzig2021role}.
Beyond these broader efforts, our own prior work introduced a reproducibility checklist for computational social science developed through a systematic literature review and survey evaluation \cite{Momeni2025ChecklistsSocialScience}. While that study focused on identifying reporting gaps and codifying best-practice recommendations, it did not examine how such documentation practices influence actual reproduction attempts. The present work addresses this gap by empirically assessing how curated documentation and execution environments affect concrete reproduction outcomes, user experience, and error patterns across diverse CSS methods.

\subsection{Positioning Our Contribution}

Our work builds on this literature in three ways. First, it brings together concerns from ML/NLP reproducibility and social data analysis by focusing on computational social science methods that combine complex code, non-trivial environments, and often sensitive data. Second, rather than evaluating checklists and documentation only indirectly through reporting quality, we study their impact on \emph{actual reproduction attempts}: we compare curated versus uncurated documentation and a preset execution environment in terms of success rates, task time, error profiles, and user experience. Third, we adopt a human-centred perspective, analysing how users with varying technical backgrounds interact with documentation, code, and platforms such as Methods Hub, and how they rely on external resources (including AI tools) to overcome remaining gaps. This allows us to move beyond abstract recommendations toward empirically grounded design implications for reproducible, usable computational methods.

 To guide the empirical analysis, we address the following research questions:

 \textbf{RQ1:} How does documentation quality affect execution-level reproducibility outcomes (success, errors, time)?

  \textbf{RQ2:} How does environment standardization influence technical barriers and assistance needs?

 \textbf{RQ3:} How do documentation and environment support affect the interpretability and users’ confidence in reproduced results?

\section{Methodology}
This section outlines the study design, participant recruitment strategy, session protocol, and measures used to evaluate the reproducibility and usability of computational methods across platforms.

\subsection{Study Design}

This study evaluates the reproducibility of NLP and machine learning workflows commonly used in computational social science (CSS). The evaluation compares two types of implementations for each workflow: (1) curated versions hosted on the Methods Hub, and (2) comparable implementations sourced from external platforms such as GitHub or Hugging Face. The study consists of two phases: an initial usability testing phase and an intervention phase that introduced a preset execution environment via myBinder.

\subsection{ Computational Methods and Repositories}

To prepare the usability study, our team curated a set of ten computational methods—primarily NLP and machine learning workflows—that are commonly used in computational social science (CSS) for tasks such as text retrieval, topic modeling, bias assessment, sentiment analysis, and digital trace data collection. For each curated method, we identified a comparable implementation hosted externally on GitHub or Hugging Face, allowing us to evaluate reproducibility across two types of platforms: the Methods Hub and widely used public code repositories. This resulted in ten method pairs (one Methods Hub version and one external version), yielding a total of 20 repositories used in the final usability evaluation. We validated the reproducibility of all methods and ensured that every implementation runs reliably from start to finish.

Participants tested the following ten computational methods frequently applied in CSS research:

\begin{itemize}
\item Claims Retrieval
\item Semantic Search over Social Media Posts
\item Multilingual Text Detoxification (mDetoxifier)
\item Discovering Themes with Topic Modeling
\item SSciBERT Tutorial (Politics Domain)
\item Propensity Score Matching
\item 4Chan Data Collection
\item Text Edit Distance
\item Word Embeddings Bias Analysis
\item Topic-Based Sentiment Analysis
\end{itemize}

\subsection{Participants}
 Participants were recruited from a diverse set of disciplinary backgrounds, including social sciences, computer science, and related fields. The study did not restrict participation to any particular user type; instead, the only inclusion criterion was that participants possessed a basic familiarity with GitHub and the ability to run R- or Python-based scripts. This ensured that all participants could meaningfully attempt the reproduction tasks while still representing a broad range of experience levels.

Although participants varied substantially in academic discipline, coding experience, and repository familiarity, these characteristics were not used to assign tasks or define experimental groups. All participants were randomly assigned a computational method to reproduce. Their background variables—discipline, programming experience, and GitHub familiarity—were collected during screening and later used in correlation analyses to examine relationships with task success, error rates, and assistance frequency.

Recruitment was carried out using an online screening, which captured disciplinary affiliation and technical experience. This allowed us to build a participant pool reflecting the diversity of real-world users of computational research methods, including both novice social scientists and technically proficient researchers.
\subsection{Procedure}
Participants were randomly assigned to one of the available repositories, which were pre-classified into the three reproduction conditions (A–C). Each participant completed only one reproduction task. Because repositories structurally differed across conditions (curated vs. uncurated; manual vs. preset environment), assignment occurred within these predefined condition pools rather than through full random allocation across identical interventions. Thus, while random assignment was used to reduce selection bias, the study does not constitute a Randomized Controlled Trial with matched tasks across conditions. Participants were recruited to attempt method reproduction tasks. Each participant was assigned a single computational method to reproduce, drawn either from the Methods Hub or from an external repository. Sessions were recorded with participant consent, and participants were encouraged to think aloud. They were allowed to use external help resources such as Google or ChatGPT.

\subsubsection{Reproduction Conditions}

The study compared method reproducibility across three distinct reproduction conditions, which differed in documentation quality and execution environment.

\paragraph{Condition A: Uncurated Documentation + Manual Environment Setup}
Participants reproduced methods sourced from public external repositories (GitHub or Hugging Face). These repositories were not curated by our team and contained their original documentation. A total of 18 usability tests were conducted under this condition (13 GitHub, 5 Hugging Face).

\paragraph{Condition B: Curated Documentation + Manual Environment Setup}
19 usability tests were conducted using repositories curated by our team according to the Methods Hub checklist guidelines\footnote{https://github.com/GESIS-Methods-Hub/guidelines-for-methods/blob/main/README-template.md}. These repositories included enhanced documentation and clearer execution instructions but required users to set up the environment manually.

\paragraph{Condition C: Curated Documentation + Preset Execution Environment}
To reduce friction from installation and versioning issues, we introduced a preset myBinder\footnote{https://mybinder.org/} execution environment for the curated repositories. Ten usability tests were conducted under this condition.

Across all three reproduction conditions, a total of 47 usability tests were completed.

\subsubsection{Session Structure}
Each usability test session followed a four-tier structure:

\paragraph{Tier 1: Introduction and Consent}
Participants provided consent, were introduced to the study goals, and were reminded that they could freely use external help resources.

\paragraph{Tier 2: Method Introduction}
Moderators introduced the assigned method using contextual examples and visual aids (e.g., FigJam demos).

\paragraph{Tier 3: Reproduction Task}
Participants attempted to execute the workflow—primarily in Google Colab to standardize environments, though local execution was permitted. Moderators observed silently (fly-on-the-wall) while participants verbalized thoughts (think-aloud protocol).

\paragraph{Tier 4: Reflection and Post-Test Survey}
Participants reflected on their experience and completed a structured feedback survey.\footnote{\url{https://forms.gle/Nhe7wseNEjhibuMH8}}

This structured progression enabled consistent comparison across participants and platforms.

\subsection{Measures}
To operationalize reproducibility in computational social science workflows, we adopted a multi-dimensional approach. Reproducibility is defined here as a user’s ability to execute a method, obtain outputs, and understand the procedural and conceptual steps involved. We therefore combined behavioral indicators (task completion, time on task, error types, assistance frequency) with post-task questionnaire measures of clarity, interpretability, and perceived correctness.
This combination allows us to capture both execution-level reproducibility (whether the workflow runs successfully) and user-level interpretability (whether users can understand and evaluate the outputs in practice). By integrating objective performance metrics with subjective assessments, we obtain a more comprehensive understanding of the technical and conceptual barriers that shape reproducibility outcomes in computational social science.

\subsubsection{Quantitative Measures}
The usability study was guided by a set of qualitative variables and measures that allowed us to capture not only the observable outcomes of user interaction but also the experiential dimensions of working with computational methods. These were defined as follows:
\begin{itemize}
    \item \textbf{Task Completion (Reproducibility).} Whether the workflow was successfully reproduced.  
    \item \textbf{Time on Task.} Total time required to complete (or fail to complete) the workflow. 
    \item \textbf{Error Rates.} Number and type of Code A, B, and C errors.  
    \item \textbf{Assistance Frequency.} Number of times participants sought external help.
\end{itemize}
Results are visualized in Figures~\ref{fig:success_rate_fig}, \ref{fig:error_distribution_fig}, and Table ~\ref{tab:assistance_frequency}.

\paragraph{Error Coding}
Errors encountered during reproduction were categorized into three types:
\begin{itemize}
  \item \textbf{Code A: Participant errors.} E.g., incorrect order of execution, syntax errors introduced by the participant.
  \item \textbf{Code B: Repository errors.} E.g., missing data, unclear documentation, deprecated libraries.
  \item \textbf{Code C: System errors.} Environment issues.
\end{itemize}
Code B errors were prioritized analytically due to their direct link to repository usability.

\subsubsection{Qualitative Measures}
\begin{itemize}
    \item \textbf{Thematic Analysis.} Post-test reflections and transcripts were coded to identify recurring challenges and strategies.
\end{itemize}

\section{Results}
This section presents the quantitative and qualitative findings from 47 usability tests conducted across the three reproduction conditions:
\begin{itemize}
    \item (A) Uncurated Documentation + Manual Setup,
    \item (B) Curated Documentation + Manual Setup, and
    \item (C) Curated Documentation + Preset Execution Environment (myBinder).
\end{itemize}

The results are structured to address our research questions concerning (1) execution-level reproducibility, (2) environment-related barriers and support needs, and (3) interpretability and output confidence. Quantitative findings are presented first, including task completion (reproducibility), time on task, error distribution, and assistance patterns. These measures capture observable performance differences across conditions. We then present qualitative analyses based on session transcripts and post-test reflections, which contextualize these patterns and provide insight into the technical and conceptual mechanisms underlying reproduction success or failure.
\subsection{Quantitative Results}
The quantitative analyses directly address execution-level reproducibility (RQ1) and environment-related barriers and assistance patterns (RQ2).

\subsubsection{Participant Background and Programming Familiarity}
We compared self-reported programming familiarity (1–5 scale) across conditions (Table~\ref{tab:familiarity_by_condition}). A Kruskal--Wallis test showed no significant difference, $H(2)=4.84$, $p=.089$. 
\begin{table}[h!]
\centering
\caption{Self-reported programming familiarity (1–5 scale) across reproduction conditions.}
\label{tab:familiarity_by_condition}
\begin{tabular}{lccc}
\toprule
\textbf{Condition} & \textbf{n} & \textbf{Mean} & \textbf{SD} \\
\midrule
A (Uncurated) & 18 & 3.44 & 1.10 \\
B (Curated, no myBinder) & 19 & 2.79 & 0.85 \\
C (Curated + myBinder) & 10 & 2.80 & 1.03 \\
\bottomrule
\end{tabular}
\end{table}
Across all participants, familiarity was not significantly correlated with task completion ($\rho=-0.11$, $p=.469$), error rate ($\rho=0.16$, $p=.277$), or time on task ($\rho=-0.17$, $p=.246$). These results suggest that programming background does not explain the observed performance differences.

\subsubsection{Task Completion (Reproducibility)}
\begin{figure}
    \centering
    \includegraphics[width=0.45\textwidth]{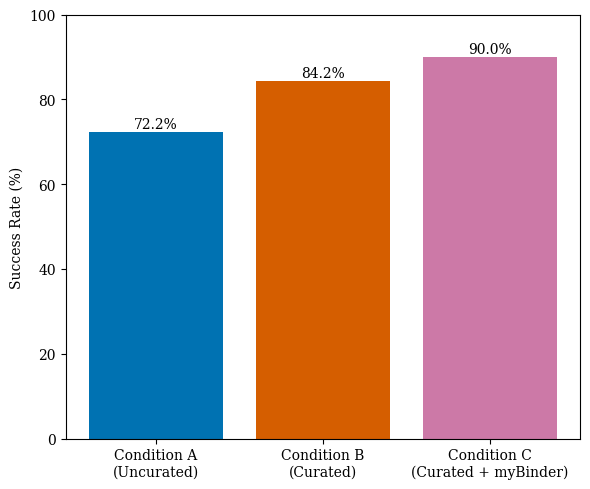}
    \caption{Task completion (reproducibility) rates across reproduction conditions (larger is better).}
    \label{fig:success_rate_fig}
\end{figure}

Task completion rates differed across the three reproduction conditions (Figure \ref{fig:success_rate_fig}).

Repositories with curated documentation (Condition B) showed a higher average success rate (84.2 \%) than uncurated external repositories (72.2 \%).
The curated repositories executed within a preset myBinder environment (Condition C) achieved the highest task completion rate at 90.0 \%.

However, a chi-square test indicated no statistically significant association between condition and task completion ($\chi^2(2) = 1.54$, $p = .462$, Cramér’s $V = .18$).
\subsubsection{Time on Task} 
\begin{figure}
    \centering
    \includegraphics[width=0.45\textwidth]{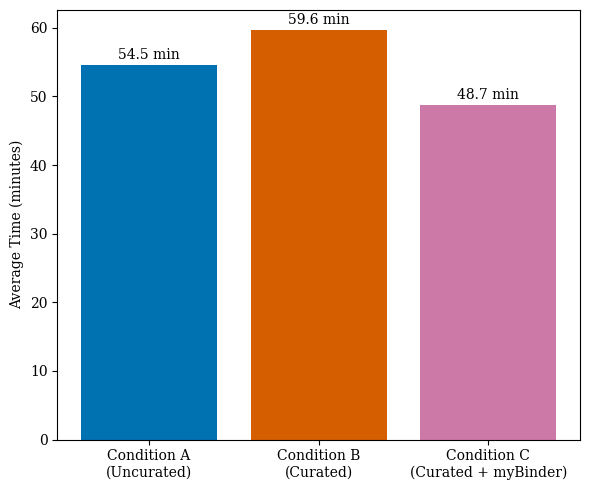}
    \caption{Average task completion time across reproduction conditions.}
    \label{fig:task_time_fig}
\end{figure}

As shown in Figure~\ref{fig:task_time_fig}, average time on task also varied across conditions.
External repositories required the least time on average (54.5 minutes), often due to earlier termination of unsuccessful attempts, followed by curated repositories without a preset environment (59.6 minutes).
The curated repositories with myBinder support (Condition C) were substantially faster, with an average completion time of 48.68 minutes.
This reduction indicates that eliminating environment-setup overhead leads to more efficient reproduction.

However, a Kruskal–Wallis test showed no statistically significant difference in task duration across conditions ($H(2) = 0.28$, $p = .870$).

\subsubsection{Error Rates}
A total of 128 errors were observed across all usability tests. Uncurated external repositories (Condition~A) accumulated the highest number of errors (65), followed by curated repositories without a preset environment (Condition~B: 55). In contrast, the curated repositories executed within a preset myBinder environment (Condition~C) drastically reduced the total number of errors to only eight.

\begin{figure}
\centering
\includegraphics[width=0.45\textwidth]{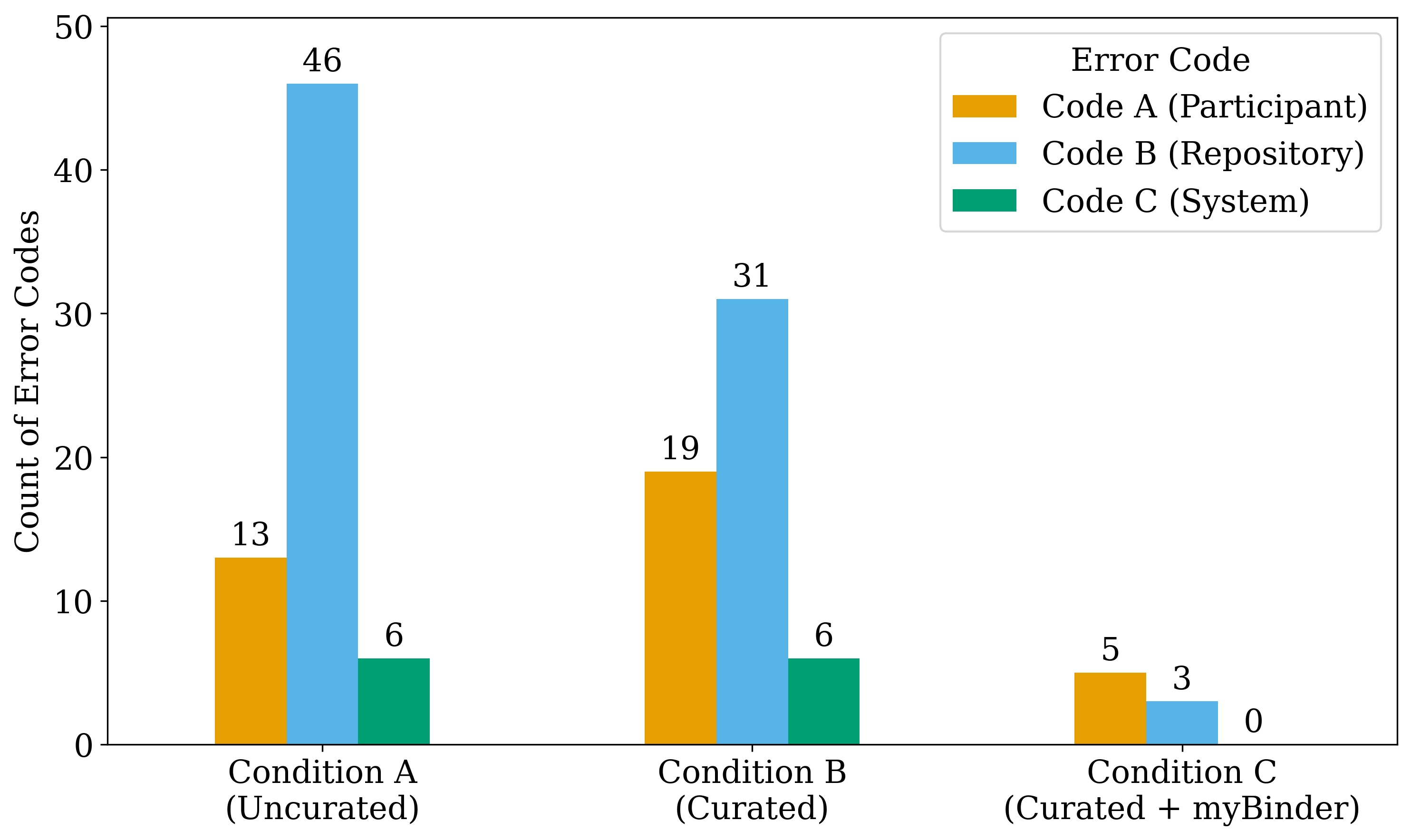}
\caption{Distribution of errors by codes across reproduction conditions (smaller is better).}
\label{fig:error_distribution_fig}
\end{figure}

Figure~\ref{fig:error_distribution_fig} shows the distribution of error types across reproduction conditions. Repository-related errors (Code~B), including missing files, broken dependencies, and inconsistent instructions, were the dominant error type in Conditions~A and~B. System-related errors (Code~C) occurred only in Conditions~A and~B and were completely eliminated under Condition~C.

While participant-related errors (Code~A) persisted across all conditions, their relative contribution increased in Condition~C, reflecting a shift from execution failures toward higher-level interaction or interpretation issues. Overall, these results indicate that environment standardization substantially mitigates technical and repository-related error sources, leaving fewer but more conceptually driven errors.

To assess statistical robustness, we conducted a Kruskal–Wallis test across the three conditions. Repository-related error rates differed significantly, $H(2) = 9.11$, $p = .010$.
Post-hoc Dunn tests with Holm correction revealed that Condition A (Uncurated) exhibited significantly higher repository-related error rates than Condition C (Curated + myBinder) ($p = .009$). Condition B (Curated without myBinder) also showed significantly higher error rates than Condition C ($p = .033$). No significant difference was observed between Conditions A and B ($p = .490$).
These results indicate that environment standardization, rather than documentation alone, played a central role in reducing execution instability.

\subsubsection{Assistance Frequency}
Table~\ref{tab:assistance_frequency} summarizes assistance requests across reproduction conditions. Uncurated repositories (Condition~A) generated the highest assistance burden, with 60 assistance requests, nearly three quarters of which (73.3\%) were caused by repository-related issues such as unclear instructions, missing files, or dependency failures. 
\begin{table}
\centering
\caption{Assistance requests across reproduction conditions. Repository-related assistance refers to requests caused by unclear instructions, missing files, dependency conflicts, or code-level issues.}
\small
\begin{tabular}{@{}l@{\hspace{18pt}}cc@{}}
\toprule
\textbf{Condition} 
& \multicolumn{2}{c@{}}{\textbf{Assistance requests}} \\
\cmidrule{2-3}
& \textbf{Total} 
& \textbf{Repository-related} \\
\midrule
\textbf{A (Uncurated)} 
& 60 & 44 (73.3\%) \\
\textbf{B (Curated)} 
& 48 & 29 (60.4\%) \\
\textbf{C (Curated + myBinder)} 
& 10 & \phantom{0}2 (20.0\%) \\
\bottomrule
\end{tabular}
\label{tab:assistance_frequency}
\end{table}
Curated documentation without environment support (Condition~B) reduced the total number of assistance requests, but repository-related issues still accounted for the majority of assistance (60.4\%). 

In contrast, the curated repositories executed within a preset environment (Condition~C) required very little assistance overall, and only 20.0\% of requests were repository-related. This sharp reduction indicates that environment standardization substantially mitigates repository fragility, shifting assistance needs away from execution failures toward general or exploratory questions.

\subsection{Qualitative Results}
A thematic analysis of the session transcripts and post-test reflections identified seven recurring usability challenges and experiential patterns. These themes contextualize and help explain the quantitative differences observed across the three reproduction conditions (see Table~\ref{tab:Qs_responses}).

To ensure analytic clarity, the themes are organized in relation to our research questions: themes concerning documentation clarity and technical barriers (Q1, Q2, Q4, and Q6) primarily address RQ1 (execution-level reproducibility), themes related to environment support and assistance patterns (Q7 and Q8) inform RQ2 (environmental friction and user support), and themes focused on output interpretation and perceived correctness (Q3 and Q5) speak to RQ3 (interpretability and conceptual reproducibility).

\begin{table}[h!]
\centering
\caption{Post-test questionnaire results across reproduction conditions (Q1--Q3 = mean Likert scores; Q4--Q8 = percentage-based responses). For Q8, “No help needed” indicates the participant completed the task independently; 
“AI/README/online help” indicates self-directed troubleshooting without human intervention; 
and “Facilitator assistance” reflects direct help from the session moderator.}
\small\setlength{\tabcolsep}{3pt}
\begin{tabular}{@{}l@{\hspace{12pt}}ccc@{}}
\toprule
& \multicolumn{3}{c}{\textbf{Condition}} \\
\cmidrule{2-4}
\textbf{Question / Response Type} 
& \textbf{A} 
& \textbf{B} 
& \textbf{C} \\
& \textbf{(Uncurated)} & \textbf{(Curated)} & \textbf{(Curated +} \\
&&& \textbf{myBinder)} \\
\midrule
\multicolumn{4}{@{}l}{\textbf{Mean Scores (1–5)}} \\
\textbf{Q1:} Instruction clarity (mean) 
& 2.94 & 3.42 & 3.40 \\
\textbf{Q2:} Setup clarity (mean)
& 2.83 & 2.95 & 2.70 \\
\textbf{Q3:} Output correctness (mean)
& 2.83 & 3.58 & 3.10 \\
\midrule
\multicolumn{4}{@{}l}{\textbf{Percentage-Based Responses}} \\
\textbf{Q4:} Major code edits required
& 55.6\% & 36.8\% & 0.0\% \\
\textbf{Q5:} Able to compare output
& 22.2\% & 63.2\% & 20.0\% \\
\textbf{Q6:} Steps clearly in one place
& 44.4\% & 73.7\% & 60.0\% \\
\textbf{Q7:} Beginner-friendly
& 44.4\% & 36.8\% & 60.0\% \\
\textbf{Q8:} No help needed
& 33.3\% & 21.1\% & 20.0\% \\
\hspace{0.6cm}AI/README/online help
& 44.4\% & 68.4\% & 70.0\% \\
\hspace{0.6cm}Facilitator assistance
& 22.2\% & 10.5\% & 10.0\% \\
\bottomrule
\end{tabular}
\label{tab:Qs_responses}
\end{table}

\subsubsection{Documentation and Guidance Inadequacies}
Across all conditions, inadequate documentation was a recurring barrier. Participants reported unclear distinctions between mandatory and optional steps, insufficient detail in README files, and missing guidance for environment setup and file handling. These issues were particularly challenging for less technical users.

Questionnaire results support these observations (Table~\ref{tab:post_test_results}). Instruction clarity (Q1) improved under curated documentation (Condition B) relative to the uncurated condition (A) and remained comparable in Condition C. Similarly, the proportion of participants indicating that all required steps were presented in a single place (Q6) increased under curated documentation. These findings indicate that curation reduces ambiguity, though it does not eliminate procedural complexity.

\subsubsection{Technical Implementation Barriers}
Across all reproduction conditions, participants encountered substantial technical barriers that impeded smooth execution of the methods. These issues included dependency conflicts, missing files, version mismatches, and broken or deprecated APIs. Several participants characterized the experience as “everything is an error,” reflecting repeated failures during installation or execution. R-based methods posed additional challenges for participants without prior R experience, who frequently struggled with unfamiliar package systems and error messages. Such technical obstacles often halted progress entirely or required extensive troubleshooting.

The questionnaire responses reflect these difficulties. Setup clarity (Q2), which captures the transparency of installation steps, dependency requirements, and file preparation, remained low across all conditions (Table~\ref{tab:Qs_responses}). While curated documentation (Condition~B) offered a slight improvement over the uncurated condition (A), clarity declined again when the curated methods were paired with the preset execution environment (C). Because myBinder eliminated many installation steps, lower Q2 ratings in Condition C may reflect uncertainty about the remaining instructions rather than increased setup complexity. This pattern indicates that documentation and environment standardization alone did not fully resolve the underlying technical complexity of the methods.

\subsubsection{Accessibility Challenges for Non-Technical Users}
Participants with limited programming experience—particularly those from social science backgrounds—struggled with repositories that lacked step-by-step instructions or conceptual explanations. Without narrative guidance or worked examples, several described code snippets as “random things,” underscoring the need for scaffolding that connects computational steps to methodological intent.

The questionnaire responses reflect these accessibility challenges. Ratings of beginner-friendliness (Q7) varied markedly across conditions (Table~\ref{tab:Qs_responses}). Curated documentation (Condition~B) did not substantially improve perceptions of accessibility relative to the uncurated condition (A), suggesting that clearer wording alone did not reduce the cognitive demands placed on novice users. The most pronounced improvement occurred when curated documentation was combined with a preset execution environment (Condition~C), indicating that removing environment-setup tasks—rather than adjusting documentation—played the central role in enabling novices to engage with the methods. Together, these findings highlight that non-technical users benefit most when both conceptual explanations and technical barriers are addressed.

\subsubsection{Reliability of Code Execution}
Participants noted clear differences in the technical reliability of code execution across conditions. Curated repositories generally exhibited more stable execution: file structures were intuitive, dependencies were more consistently defined, and code was often better commented, helping users understand the workflow and troubleshoot issues. Although errors still occurred, they were typically resolvable without extensive intervention. This stability was further enhanced when curated documentation was paired with a preset execution environment (Condition C).

In contrast, uncurated external repositories frequently contained outdated code, broken APIs, missing files, or unresolved dependency conflicts. These issues often prevented participants from completing the method within the session timeframe, severely limiting reproducibility. 

The questionnaire responses support these observations (Table~\ref{tab:Qs_responses}). The need for major code modifications (Q4) was highest in the uncurated condition (A) and decreased substantially under curated documentation (B). The curated documentation paired with a preset execution environment (C) offered the greatest stability, with no participants reporting major code edits. Together, these findings indicate that curation—and especially environment standardization—plays a critical role in ensuring reliable code execution.

\subsubsection{Reliance on External AI Support}
Across all conditions, participants frequently relied on external AI tools (e.g., ChatGPT, Gemini AI) for troubleshooting.  
Users working with curated repositories (Condition~B) or curated repositories executed with the preset environment (Condition~C), while frequently using AI assistance, were generally more likely to achieve successful outcomes with guided support.  
In contrast, uncurated external repositories (Condition~A) created heavier dependence on AI, with participants often requiring step-by-step guidance even for basic functionality. The inadequacy of official documentation in this condition amplified this reliance.

These patterns are reflected in the questionnaire responses (Table~\ref{tab:Qs_responses}, Q8). Across all conditions, many participants made use of AI or online resources when they encountered issues (Q8). Importantly, facilitator assistance was highest in the uncurated condition (A) and substantially lower in Conditions~B and~C. This indicates that curated documentation—and especially the standardized environment—reduced the need for human intervention. While AI-based help appeared more frequently in B and C, this reflects increased self-sufficiency rather than increased difficulty: participants were able to address issues independently rather than relying on the moderator.

\subsubsection{Result Interpretation and Output Quality}
Curated repositories (Conditions~B and C) generally produced clearer and more interpretable results. Methods such as political text classification and topic-based sentiment analysis yielded outputs that participants found meaningful, provided the inputs were properly formatted.  
In contrast, uncurated external repositories (Condition~A) often produced inconsistent or confusing results. Limited explanation of scoring metrics and missing examples left participants unsure of how to interpret outputs. For instance, similarity scores in the Claims Retrieval method were unexpectedly low, prompting questions about effectiveness and correctness of the underlying implementation.

Responses to output correctness (Q3) and the ability to compare results with a reference (Q5) align with these observations (Table~\ref{tab:Qs_responses}).  
Condition~A showed the lowest perceived correctness and the lowest ability to compare outputs, indicating unclear or ambiguous results.  
Condition~B improved substantially on both measures, suggesting that curated documentation supported interpretability.  
Condition~C maintained moderate perceived correctness but again showed low comparability, reflecting that a preset environment improves execution reliability but does not, on its own, provide the interpretive scaffolding needed for users to judge output quality.

\section{Discussion}

This study evaluates reproducibility across three conditions—uncurated documentation (A), curated documentation (B), and curated documentation with a preset execution environment (C). By combining behavioral KPIs with questionnaire and qualitative data, we examine how technical and conceptual factors shape reproduction outcomes.

Prior work consistently emphasizes that  reproducibility improves when researchers provide structured documentation (e.g., checklists, detailed reporting, README/metadata) and make the computational environment explicit and portable (e.g., via dependency specification and environment encapsulation) \cite{pineau2021improving,magnusson2023reproducibility,leipzig2021role,albertoni2023reproducibility,semmelrock2025reproducibility,collberg2015repeatability,obels2020analysis,hardwicke2021analytic,clyburne2019computational}. Our findings are consistent with this broader literature. However, unlike audit-based or policy-oriented studies, this work provides a controlled, user-centered comparison of curated and uncurated repositories, separating the effects of documentation and execution environments while measuring behavioral performance (completion, time, error types, and assistance). This experimental design allows us to quantify how different layers of infrastructure influence reproducibility in practice.

\subsection{Reproducibility Performance: Success Rates, Time, and Error Profiles}

Descriptively, success rates increased from 72.2\% (A) to 84.2\% (B) and 90\% (C), though differences were not statistically significant.

The success rate for Condition~C did not reach 100\% due to a memory (RAM) limitation in the myBinder environment: one method exceeded the available resources and terminated unexpectedly. This failure was unrelated to documentation quality or code correctness and instead reflects infrastructural constraints common in cloud-based execution environments.

Task time showed a similar descriptive trend, with Condition C fastest (48.7 minutes), though differences were not statistically significant.

Error analysis provides deeper insight into these performance differences. Condition~A accumulated the highest number of errors (65), with 46 (70.7\%) attributable to repository-level issues (Code~B), such as missing files, broken dependencies, and ambiguous instructions. In contrast, Condition~B exhibited fewer total errors (55), with a notable reduction in Code~B errors (30). Condition~C further reduced total errors to eight, eliminating all system-level failures (Code~C) and nearly all repository-level issues.

Together, these results indicate that reproducibility failures in uncurated repositories are strongly associated with technical fragility and unstable execution environments. While curated documentation reduced ambiguity and improved workflow clarity, statistically significant reductions in repository-level errors were observed primarily when documentation was combined with environment standardization (Condition~C). The preset environment removes nearly all execution-related barriers but cannot compensate for missing conceptual explanations or absent reference outputs.

\subsection{Documentation Quality and Conceptual Clarity as Determinants of Interpretability}

Curated documentation played a central role not only in improving execution success but also in shaping users’ ability to interpret method outputs. Perceived correctness (Q3) increased from 2.83 under Condition~A to 3.58 under Condition~B, reflecting clearer output behavior and more coherent workflow descriptions. The ability to compare results with reference outputs (Q5) rose sharply from 22.2\% in Condition~A to 63.2\% in Condition~B, underscoring the importance of examples and clearly defined metrics.

Condition~C, despite delivering the highest execution reliability, showed a substantially lower ability to compare outputs (20.0\%). This decline does not indicate reduced interpretability caused by the environment but reflects that several methods lacked reference outputs or sufficiently explained evaluation metrics. \textbf{Environment standardization ensures that code runs, but interpretability depends on the presence of conceptual scaffolding—including reference outputs, metric descriptions, and usage examples.}

In other words, execution success is necessary but not sufficient for meaningful reproducibility. Without interpretive support, even flawlessly executed methods sometimes leave users uncertain about the correctness of the reproduction.

This study primarily evaluates execution-level reproducibility. When reference outputs were available, participants compared their results to expected outputs, enabling partial output validation. We did not independently verify equivalence with original published claims; therefore, our conclusions focus on executability and user-level comparability rather than full scientific replication.
\subsection{Implications for Reproducibility Infrastructure}

The combined findings from KPIs, error analysis, and qualitative feedback suggest several implications for the design of platforms supporting reproducibility in computational social science:

\begin{itemize}
    \item \textbf{Documentation curation reduces ambiguity.} and improves interpretability, though statistically significant error reductions were strongest when paired with environment standardization. 
    \item \textbf{Environment standardization addresses system-level fragility.} Condition~C virtually eliminated environment-related failures, though its performance was ultimately constrained by external RAM limitations, illustrating that executability is still bounded by infrastructural resources.
    \item \textbf{Interpretability remains a critical gap.} Missing reference outputs and unclear metric explanations were major barriers across all conditions, especially in Condition~C.
    \item \textbf{AI-assisted troubleshooting complements documentation.} Participants frequently used third-party AI tools to resolve issues, suggesting that future documentation practices and platforms may benefit from considering AI-supported troubleshooting as part of typical user workflows.
\end{itemize}

Overall, the study demonstrates that reproducibility barriers are multi-layered: technical, procedural, and conceptual obstacles interact and compound one another. Effective reproducibility infrastructure must therefore integrate curated documentation, stable execution environments, and clear interpretive guidance. Only by addressing all three dimensions can platforms fully support reliable and transparent computational research workflows.
\section{Limitations}

This study has several limitations that should be considered when interpreting the findings.

First, while we contrast curated documentation with uncurated external repositories, we did not systematically assess or quantify the baseline quality of documentation in the methods. Consequently, we might have drawn a biased sample (e.g., methods with a higher documentation quality), which may have influenced our results.

Second, methods compared across conditions were not formally matched on computational complexity, dependency structure, implementation effort, or interpretability demands. For each curated Methods Hub workflow, we identified a comparable external repository addressing the similar analytical task based on realistic user search behavior. While this design reflects how researchers encounter tools in practice, inherent differences across publicly available implementations may have influenced performance independently of documentation or environment support. Accordingly, our findings should be interpreted as evidence of reproducibility differences under realistic workflow conditions rather than as strictly controlled causal estimates. Future work could employ tightly matched benchmark tasks to isolate intervention effects more precisely.

Third, participants varied substantially in programming familiarity, technical background, and disciplinary training. These differences shaped how users perceived method complexity, documentation clarity, and reproducibility barriers, introducing additional variability that could not be fully controlled.

Finally, this study focuses on execution-level reproducibility, that is, whether users can run methods and obtain outputs, rather than on conceptual replication or scientific validity.

\section{Conclusion}

This study examined how documentation quality and execution environment shape the reproducibility of computational social science methods. By comparing three reproduction conditions—uncurated documentation (Condition~A), curated documentation (Condition~B), and curated documentation paired with a preset execution environment (Condition~C)—we identified the distinct technical and conceptual barriers that users encounter during method reproduction. By integrating behavioral KPIs, error diagnostics, questionnaires, and thematic analyses, we provide a multi-layered understanding of reproducibility barriers.

Our findings show that curated documentation improves workflow clarity and users’ ability to interpret method outputs. Statistically significant reductions in repository-level errors were observed primarily when curated documentation was paired with a standardized execution environment. However, environment standardization alone is insufficient: methods remained difficult to interpret when reference outputs or metric explanations were incomplete or missing. This underscores that reproducibility is not only a technical but also a conceptual property.

The study provides empirical observations of how AI-assisted troubleshooting is currently employed in reproduction workflows. Participants frequently relied on AI like ChatGPT to resolve ambiguities or execution issues, suggesting that current reproduction workflows increasingly blend documentation, platform features, and AI-driven support. Importantly, curated documentation and a standardized environment reduced the need for facilitator intervention, enabling users to resolve issues on their own.

Overall, reproducibility barriers were found to be multi-layered, spanning documentation quality, environment stability, and conceptual clarity. Addressing these dimensions in combination—rather than isolation—is essential for building reliable and accessible reproducibility infrastructures for computational social science. Platforms that integrate high-quality documentation, stable execution environments, reference outputs, and AI-compatible workflows are likely to offer the strongest support for transparent and verifiable computational research.

\section*{Data and Code Availability}

All materials, including the survey instruments, anonymized response data, and checklist resources, are published in the associated public repository \footnote{https://github.com/momenifi/methods-hub-reproducibility-study}.

\begin{acks}
This research was supported by TIER2 (Grant No. 101094817), with additional support from GESIS – Leibniz Institute for the Social Sciences through the Digital Behavioral Data project.
\end{acks}

\bibliographystyle{ACM-Reference-Format}
\bibliography{sample-base}

\appendix


\end{document}